\documentclass{svproc}

\usepackage{graphicx}
\usepackage{multicol}
\usepackage{footmisc}
\usepackage{hyperref}
\usepackage{nameref}
\usepackage{amsmath}
\usepackage{cite}

\usepackage{xcolor}

\newcommand\new[1]{\textcolor{black}{#1}}  
\newcommand\added[1]{\new{#1}}
\usepackage{soul}
\newcommand\deleted[1]{}
\newcommand\replace[2]{\deleted{#1}\new{#2}}
\newcommand\reviewcomment[1]{}

\begin{document}
\mainmatter

\title{Resilience Against Soft Faults through Adaptivity in Spectral Deferred Correction}
\titlerunning{Resilience in SDC}


\author{Thomas Saupe\inst{1, 2} \and Sebastian G{\"o}tschel\inst{2} \and Thibaut Lunet\inst{2} \and Daniel Ruprecht\inst{2} \and Robert Speck\inst{1}}
\authorrunning{Saupe et al.}


\institute{J{\"u}lich Supercomputing Centre, Wilhelm-Johnen-Stra{\ss}e, \\52428 J{\"u}lich, Germany \and Hamburg University of Technology, Am Schwarzenberg-Campus 3, 21073 Hamburg, Germany}
%




\maketitle

\begin{abstract}
As supercomputers grow in hardware complexity, their susceptibility to faults increases and measures need to be taken to ensure the correctness of results. Some numerical algorithms have certain characteristics that allow them to recover from some types of faults.
It has been demonstrated that adaptive Runge-Kutta methods provide resilience against transient faults without adding computational cost.
Using recent advances in adaptive step size selection for spectral deferred correction (SDC), an iterative numerical time stepping scheme that can produce methods of arbitrary order, we show that adaptive SDC can also detect and correct transient faults. Its performance is found to be comparable to that of the dedicated resilience strategy Hot Rod.
\end{abstract}

\section{Introduction}\label{secIntroduction}
While computers have the reputation of performing arithmetic without errors in the general public, this notion is rather new.
A prominent example of the poor reputation of electronic computation well into the sixties features John Glenn, the first American to perform a full rotation around the Earth in space.
He famously trusted crucial flight information computed by an electronic computer only after it was confirmed by one of NASA's legacy human computers equipped only with a slide rule~\cite{johnson_reaching_2019}.
In the early days of computing, the vacuum tubes that provided the basis for implementations of logic in hardware were prone to failure and quite unreliable~\cite{miller_chip_2022}.
In particular, they shared many problems with light bulbs, notably burning out, as well as attracting moths, requiring them to be regularly ``debugged''~\cite{reid_chip_2001}.

The replacement of vacuum tubes by semiconducting transistors, following their invention in the late forties, promised to alleviate reliability concerns.
However, a new problem manifested itself in the ``tyranny of numbers'', which was the term given to the issue that engineers found themselves unable to manufacture the complex circuits that they designed on paper.
Blueprints for computers required millions of small components which had to be assembled by hand in a very error-prone process.
Computer chips became reasonably reliable only after the invention of the integrated circuit in the late fifties.
Rather than assembling each component separately, the components are essentially printed on a monolithic block of silicon.

Today, as the trend in HPC systems is for components to keep shrinking and their numbers to keep rising, the probability of a fault somewhere in the system is, again, increasing and a major concern for exa-scale systems~\cite{Dagstuhl}.
Failure of a single hardware component can cause the entire system to fail.
As an example, a faulty diode in an integrated circuit in the launch vehicle of the Mariner 8 mission resulted in a failed launch.
Instead of becoming the first spacecraft to orbit another planet and mapping out Mars, the probe ended up on the bottom of the ocean~\cite{johnson_sirens_2021}.

Faults can originate from a variety of sources such as damage to hardware components~\cite{googleFaults} or electromagnetic radiation~\cite{LAPINSKY2006267}, which has led to hospital and airplane modes on portable consumer devices.
Real-world studies suggest that faults occur multiple times per day in modern HPC centers~\cite{googleFaults},
in particular also in caches that are insufficiently protected by hardware error correction codes~\cite{FaultsRealWorldLLNL}.

Resilience against faults can be achieved via hardware and software.
Dedicated software resilience strategies often come with significant overhead, which exacerbates the computational efficiency that can be gained by adaptive methods in comparison.
The potential for adaptive time stepping schemes to improve resilience as well as computational efficiency has already been explored for embedded Runge-Kutta methods (RKMs)~\cite{embeddedRKResilience} as well as spectral deferred correction (SDC) with adaptive iteration number~\cite{grout_achieving_2017}.
\added{Adaptive methods generally measure accuracy and trigger additional work if the accuracy was found to be insufficient.
While these mechanisms are primarily developed to react to the behaviour of the problem, some transient soft faults can be corrected in adaptive methods after they have been detected in the accuracy measurement.}

SDC is a time stepping scheme that iteratively solves \added{the equations for the stages of} fully implicit RKMs \added{with preconditioned Picard iterations}.
It can be advantageous \added{in terms of computational efficiency} over the more common diagonally implicit RKMs for stiff or complex problems~\cite{doi:10.1137/18M1235405,p_adaptivity_phase_field}, when employing complex splitting techniques~\cite{FastWaveSlowWave,LAYTON2004697,minion_semi-implicit_2003} or as a basis for parallel-in-time (PinT) methods~\cite{PFASST,parallelSDC,GT_precons,parallel_sdc_hpc}.
PinT~\cite{50yrsPinT} codes especially require attention to resilience because they are designed to run on large portions of modern HPC systems.
After adaptive step-size selection for SDC has been shown to boost computational efficiency~\cite{saupe2024adaptivetimestepselection}, we now demonstrate its capacity to provide resilience against soft faults on par with a dedicated resilience strategy.
We do this by performing numerical experiments where we manually insert faults and compute recovery rates for different resilience strategies.

\section{Methods and Background\label{sec:methods}}
We give a brief introduction to SDC, including recent developments introducing step size adaptivity.
Then we will discuss different strategies for dealing with faults and explain how we design numerical experiments to test them.

\subsection{Spectral Deferred Correction}
Spectral deferred correction (SDC)~\cite{DuttSDCOriginal} is used to solve initial value problems
\begin{equation}
  \label{eq:IVP}
  u_t = f(u), \quad u(t=0) = u_0,
\end{equation}
where $u(t)$ is the solution at time $t$, the subscript \replace{$t$}{$(\cdot)_t$} marks a derivative with respect to time and $f(u)$ describes the temporal evolution of $u$.
Note that we restrict this discussion to autonomous scalar problems for the sake of simpler notation, but the method can be applied to partial differential equations (PDEs) and non-autonomous problems as well.

\autoref{eq:IVP} can be integrated numerically by applying quadrature to the right hand side $f(u)$.
This involves choosing a set of $M$ collocation nodes $\tau_m$ within the time step.
Quadrature weights $q_{mj}$ are obtained by computing the integrals of the accompanying set of interpolating Lagrange polynomials~\cite{Interpolation}.
The numerical solution $u_m$ at $t=\tau_m$ can be approximated by solving the collocation problem
\begin{equation}
  \label{eq:collocation_problem}
  u(\tau_m) \approx u_m = u_0 + \sum_{j=1}^M q_{mj} f(u_j)\,.
\end{equation}

The accuracy of the numerical integration depends on the choice of collocation nodes.
Special choices such as Gau{\ss}-Radau nodes give order $2M-1$ accuracy and are called spectral quadrature rules~\cite{hairer_wanner_II}.
However, typical quadrature rules give rise to dense matrices in the Butcher tableau and computing the solution to \autoref{eq:collocation_problem} directly can be very expensive.

Instead, SDC uses a preconditioned Picard iteration to solve~\eqref{eq:collocation_problem} iteratively.
Introducing an iteration index $k$, the SDC iteration reads
\begin{equation}
  \label{eq:SDC_iteration}
  (1 - \tilde{q}_{mm}f) \left(u_{m}^{k+1}\right) = u_0 + \sum_{j=1}^{m-1} \tilde{q}_{mj}f\left(u_j^{k+1}\right) + \sum_{j=1}^M \left(q_{mj} - \tilde{q}_{mj}\right)f\left(u_j^k\right),
\end{equation}
where $\tilde{q}_{mj}$ is the preconditioner.
The matrix containing the $\tilde{q}_{mj}$ is called $Q_\Delta$ in the literature and is \deleted{usually chosen to be}lower triangular\replace{so that it can be inverted}{, allowing inversion} by forward substitution.

The choice of preconditioner is critically important for the efficiency of SDC.
A popular choice for stiff problems is LU decomposition of the quadrature rule~\cite{LU}.
However, it is also possible to parallelize SDC by choosing diagonal preconditioners and updating the solution at all collocation nodes concurrently in each iteration~\cite{parallelSDC}.
Good preconditioners of this type have been developed for both stiff and non-stiff problems~\cite{GT_precons}, which can yield high parallel efficiency~\cite{parallel_sdc_hpc}.
It is also possible to use multiple preconditioners to realize operator splitting, such as implicit-explicit (IMEX)~\cite{FastWaveSlowWave}.

\subsection{Adaptive step size selection in SDC}
\label{subsec:adaptive_time-stepping}
The collocation problem \replace{corresponds to the equation for the stages of}{that is solved in SDC can be viewed as} a fully implicit RKM.
SDC with fixed number of iterations can also be interpreted as an RKM itself.
This suggests that it should be possible to adopt established methods from general RKM to the special case of SDC.
It has been shown that efficient step size adaptive algorithms can be constructed for SDC along the lines of embedded RKM~\cite{saupe2024adaptivetimestepselection}.

The two key ingredients are an estimate of the local error for the current step size, and an estimate of the optimal step size that would have resulted in the target local error.
In embedded RKM~\cite{Hairer_Wanner}, the error is estimated by computing two solutions of orders $p$ and $q>p$ and subtracting them to get an estimate \added{of the error} of the lower order method
\begin{align}
\label{eq:error_estimate}
    \begin{split}
    \epsilon &= \|u^{(p)} - u^{(q)}\|_{\infty} \\
    &= \|(u^{(p)} - u^{*})-(u^{(q)} - u^{*})\|_{\infty} \\
    &= \|\delta^{(p)} - \delta^{(q)}\|_{\infty} = \|\delta^{(p)}\|_\infty + \mathcal{O}(\Delta t^{q+1}),
    \end{split}
\end{align}
where $u^{(p)}$, $u^{(q)}$ are the numerical solutions obtained by integration schemes of orders $p$ and $q$.
$u^*$ marks the exact solution, $\delta$ denotes the local error with analogous meaning of the superscript and $\epsilon$ is the estimate of the local error.

The optimal step size $\Delta t_\mathrm{opt}$ to produce the desired accuracy $\epsilon\approx\epsilon_\mathrm{TOL}$ can then be inferred from the order $p$ of the error estimate, which dictates how the accuracy scales with step size $\Delta t$ using
\begin{equation}
     \Delta t_\mathrm{opt} = \beta \Delta t \left(\frac{\epsilon_\mathrm{TOL}}{\epsilon}\right)^{1/(p+1)},
     \label{eq:step_size_update}
\end{equation}
where $\beta$ is a safety factor that is usually set to $\beta=0.9$~\cite{Hairer_Wanner}.
The safety factor is important because the current step is restarted when $\epsilon > \epsilon_\mathrm{TOL}$.

In SDC, an error estimate for fixed number of iterations can be constructed from the increment, as the order increases by one with each iteration.
The resulting scheme is called $\Delta t$-adaptivity~\cite{saupe2024adaptivetimestepselection}.
It is also possible to choose both step size and iteration number adaptively in a scheme called $\Delta t$-$k$-adaptivity~\cite{saupe2024adaptivetimestepselection}.
\replace{Iteration number is chosen by declaring convergence of the collocation problem based on the residual or the increment}{The iteration number is chosen by stopping when the residual of the collocation problem or the increment reach sufficiently small values}.
Then, the error for step size adaptivity is estimated by comparing the order $M$ solution at the second to last collocation node $u_{M-1}$ to an order $M-1$ interpolation to $t=\tau_{M-1}$ of the polynomial defined by $\{u_i: i = 0, 1, ..., M-2, M\}$.

\subsection{Resilience}
\label{sec:ResilienceIntro}
Faults can occur in a variety of forms.
Damage to hardware, commonly referred to as a ``hard fault'', persists until the hardware is repaired.
On the other hand, so called ``soft faults'' are one-time events and do not occur again when repeating an operation on the same hardware.
A particularly heinous type of fault is silent data corruption\footnote{SDC is commonly used as an acronym for silent data corruption. In this article, however, SDC exclusively abbreviates spectral deferred correction.}.
This term is used to describe alterations of the solution that are not immediately apparent because the solution still seems plausible and no fault was detected during run time.
The often unknown probability of such a fault occurring can lower trust in the computational scientific method.
In this article, we investigate resilience against soft faults with the particular goal of preventing silent data corruption.

Various resilience strategies exist across hardware and software.
In hardware, error correction codes (ECCs) are standard practice and work by redundancy.
However, this comes at a cost in terms of memory footprint and energy consumption, which is why caches in processors are often equipped with ECCs that cannot correct faults~\cite{FaultsRealWorldLLNL}.
In software, replication has also been successfully employed as a generic resilience strategy~\cite{Replication}.
However, it is expensive and may not be viable on exa-scale machines as memory speeds increase insufficiently~\cite{lossyCompressionUseCases}, to the point that writing a checkpoint may take longer than the average time between faults.
These issues lead to the development of algorithm-based fault-tolerance (ABFT) strategies.
They can be tailored specifically to the algorithm they are designed to protect and have the potential to increase resilience of simulations at little extra computational cost.

\paragraph{Adaptive SDC as a resilience strategy.}
The potential of adaptive SDC to improve resilience depends on its ability to detect and recover from faults.
The \replace{SDC residual}{residual of the collocation problem} should \replace{indicate}{be increased by} soft faults that alter the solution.
The range of faults that SDC can recover from, after detecting them by increased residual, will depend on the preconditioner and the problem.
\replace{The}{A} \added{plain} Newton solver in a non-linear problem, for instance, may not converge from a faulty initial guess.
Similarly, convergence is not always guaranteed in SDC.
Still, for a wide range of faults, $k$-adaptive SDC has been shown to provide an adequate solution in the presence of faults \added{by performing additional SDC iterations}~\cite{grout_achieving_2017}.
We will call SDC with fixed step size and adaptive iteration number ``$k$-adaptivity'' in the remainder of this article.

When choosing the step size adaptively ($\Delta t$-adaptivity), we expect to detect the fault in the estimate of the local error.
Recovery happens by rejection of steps with local error exceeding the target.
This is already an effective resilience strategy for embedded RKM~\cite{embeddedRKResilience}.
$\Delta t$-$k$-adaptivity, in turn, can detect faults \replace{in}{using} the residual as well as \deleted{in}the local error estimate and can recover by means of restart if SDC does not converge after continued iteration.
Note that the tolerance determining the accuracy of adaptive SDC is a natural tolerance for fault detection.
If the fault is insignificant, it is not necessary to expend resources on recovery.
Adaptive SDC will only cause additional iterations or restarts if the fault perturbs the solution above the general accuracy requirement.

\paragraph{Hot Rod.}
To set a baseline, we compare against the dedicated resilience strategy Hot Rod~\cite{HotRod}, which was developed for explicit RKM, but can be adapted to SDC.
It is a detector for soft faults that can be tuned to high accuracy and faults can be corrected by restarting.
Detection works by computing two estimates \added{$\epsilon_1$ and $\epsilon_2$} of the local error and taking the difference to get \replace{an even more accurate estimate}{a quantity $\Delta$} that is very sensitive to faults\deleted{.}
\added{
\begin{equation}
    \Delta = \left|\epsilon_1 - \epsilon_2 \right|_\infty = \left|\delta^{(p)} + \mathcal{O}(\Delta t^q) - \delta^{(p)} + \mathcal{O}(\Delta t^r) \right|_\infty = \mathcal{O}(\Delta t^{p+2}),
    \label{eq:HR_Delta}
\end{equation}
with $q, r>p+1$.
}
For \replace{the first estimate}{$\epsilon_1$}, we use the increment in SDC.
\replace{The second estimate computes}{For $\epsilon_2$, we compute} a secondary solution at the current time \replace{by extrapolating from previous times using a Taylor method}{via a linear multistep method (LMM)~\cite[Chapter III.1]{Hairer_Wanner}, which uses the solution values from previous timesteps,} and compare to the solution obtained by SDC\added{, taking the accumulation of local errors within the LMM into account}.
The original publication \added{of Hot Rod~\cite{HotRod}} includes \replace{a Taylor method}{the error estimate~\cite[Equation 1.3]{BUTCHER1993203}} with sufficient order for \added{the RKM} Cash-Karp's method\added{~\cite{cash_carp}}\replace{.}{:}
\added{
\begin{align}
    \begin{split}
    \epsilon_2^{CK} &= \frac{1}{30}\left(u_{n-3} + 18u_{n-2} - 9u_{n-1} - 10u_n\right) + \frac{\Delta t}{10} \left(3f_{n-2} + 6f_{n-1} + f_n\right)\\
     &= \delta^{(4)} + \mathcal{O}(\Delta t^6),
    \end{split}
    \label{eq:error_estimate_HR}
\end{align}
where $u_n$ is the solution to the current timestep, $u_{n-i}$ is the solution to the timestep $i$ steps prior, and $f_{n-i}=f(u_{n-i})$.
}
We compute \replace{Taylor methods}{the coefficients of error estimates for $\epsilon_2$ as in \autoref{eq:error_estimate_HR}} at runtime in order to facilitate the easy tuning of order that is one of the advantages of SDC.

\replace{Note that both error estimates need to compute the local error of the lower order method.}{Since the increment is an estimate of the error of the second to last iterate in SDC, that is $\epsilon_1=\delta^{(k_\mathrm{max}-1)}$ with $k_\mathrm{max}$ SDC iterations, $\epsilon_2$ has to also estimate the error of the second to last iterate in order to satisfy \autoref{eq:HR_Delta}.}
This means we have to advance in time also with the lower order method and use the final iteration only for computing \replace{the error}{$\epsilon_1$}.
This is in contrast to step step size adaptive schemes, which compute the local error of the lower order method, but advance with the most accurate solution available.
For good performance it is necessary to select a detection threshold such that significant faults will be corrected, but insignificant faults or false positives do not cause too much overhead.
In the original \added{Hot Rod} paper\added{~\cite{HotRod}}, thresholds for different levels of resilience requirements are obtained by means of machine learning.
In our experiments, we simply select a problem specific tolerance by (human) trial and error.

Hot Rod is very effective at detecting faults, but comes with significant overhead.
There is computational overhead because an extra iteration is required without gain in accuracy and there is memory overhead due to the many solution-size objects that are needed for the extrapolation method.

\paragraph{Fault insertion methodology.}
While a realistic fault insertion simulation would consider faults in instructions at the lowest level, we only insert faults by flipping bits in the solution.
We do this by converting to binary IEEE 754 representation, flipping the bit and then converting back to floating point representation.
Evidence suggests that this is a sensible strategy to investigate silent data corruption, while lower level fault injection mainly increases the probability of termination of the program by the operating system~\cite{high_level_fault_injection}.
Also, we restrict the discussion to transient bit flips.
This means we assume that the fault does not occur again if we repeat the operation.

A fault can be inserted in a variety of ways.
We can insert in any iteration, at any collocation node, in any space position, bit, and at any time.
We pick a single time for fault insertion up front in order to limit the number of options.
\deleted{We then insert faults at random in up to 4000 experiments for the nonlinear Schrödinger, Allen-Cahn and Rayleigh-Benard examples.
For the Lorenz attractor problem, there are only 3840 combinations for inserting faults at a fixed time because there are 5 options for iteration number, 4 choices for collocation node (including initial conditions), 3 solution components, each with 64 bits.}
\added{All strategies that we test are fifth order accurate, meaning $k_\mathrm{max}=5$ for strategies with fixed iteration number, except $k_\mathrm{max}=6$ for Hot Rod.}
Since we want the same faults for all strategies, we restrict the random insertion \replace{to faults that would target the fixed strategy in the same way}{to the first five iterations}.
\deleted{That is, when we perform a fixed number of $k_\mathrm{max}$ iterations in the fixed and $\Delta t$-adaptivity strategies, we insert faults only in the first $k_\mathrm{max}$ iterations, even if the $k$- and $\Delta t$-$k$-adaptivity strategies may perform more iterations in this step.}
\added{
We then insert faults at random in up to 4000 experiments for PDE examples.
We also test one ordinary differential equation example, for which there are only 3840 combinations for inserting faults in the solution variables at a fixed time because there are 5 options for iteration number, 4 choices for collocation node (including initial conditions), 3 solution components, each with 64 bits.
}

\paragraph{Determining recovery rate.}
To decide if \replace{a fault is recovered}{the simulation has recovered from the fault}, we compute the global error \added{at the end of the simulation} and accept \replace{it as recovered}{successful recovery} if it is \replace{no greater than 10\,\% larger compared to the global error in}{not increased beyond some relative threshold compared to} a fault-free run.
Since the solution is not arbitrarily accurate in any case, slightly larger error is usually acceptable.
The recovery rate is then the ratio of experiments where the fault was recovered to the total number of experiments.

\subsection{Benchmark Problems}\label{sec:Problems}
\reviewcomment{Moved in-depth introduction of the problems to an appendix.}
\added{We employ four non-linear test problems in numerical experiments, which we describe in detail in Appendix~\ref{appendix:Problems}.}
\added{We use the Lorenz attractor (Appendix~\ref{sec:LorenzProblem}) as a toy problem that is very sensitive to faults and allows to study the resilience properties in detail.}
\added{Then, we test the resilience capabilities for challenging PDEs, discretized in space with spectral discretizations. 
The PDEs are the non-linear Schr{\"o}dinger equation (Appendix~\ref{sec:SchroedingerProblem}), the Allen-Cahn reaction-diffusion equation (Appendix~\ref{sec:allen_cahn_problem}), and the incompressible fluid dynamics benchmark Rayleigh-Benard convection (Appendix~\ref{sec:RBC_problem}).}
\deleted{We now introduce the four test problems that we use in numerical experiments.
All of them are non-linear.
One is an ordinary differential equation and three are partial differential equations.
We denote the simulation time at which we insert faults as $t_\mathrm{fault}$.}
All implementations are publicly available on GitHub\footnote{See \url{https://github.com/Parallel-in-Time/pySDC} for the repository} as part of the pySDC library~\cite{pySDC} and continuously tested.
Numerical experiments are run on the JUSUF supercomputer at Forschungszentrum J\"ulich~\cite{JUSUF}.

\section{Numerical Results}
\label{sec:Results}
\begin{figure}
    \centering
    \includegraphics{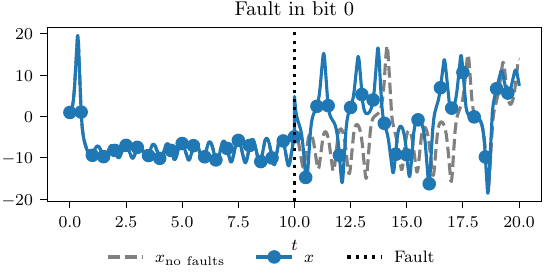}
    \caption{Horizontal coordinate of the solution of the Lorenz problem with bit 0 flipped in $x$.
    The dashed line is the solution in absence of faults, while the solid line shows the response of the fixed strategy to the fault.
    Bit 0 stores the sign, which is flipped following the fault. This dramatically changes the dynamics.}
    \label{fig:fault_bit_0}
\end{figure}

\begin{figure}
    \centering
    \includegraphics{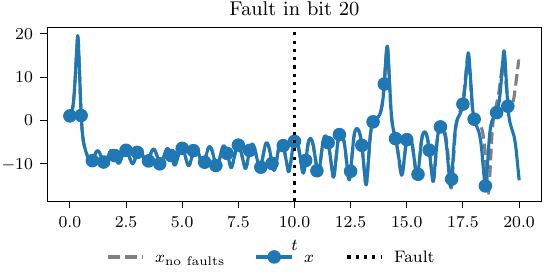}
    \caption{Horizontal coordinate of the solution of the Lorenz problem with bit 20 flipped in $x$.
    The legend is shared with \autoref{fig:fault_bit_0}.
    Bit 20 is the eighth bit in the mantissa, meaning it changes the solution only a little.
    The impact of the fault is invisible to the naked eye in the beginning, but the chaotic nature of the problem amplifies it, such that the solutions are significantly different at the end of the interval.}
    \label{fig:fault_bit_20}
\end{figure}

We test the resilience strategies discussed in \autoref{sec:ResilienceIntro} for the four benchmarks described \replace{above}{in Appendix~\ref{appendix:Problems}}.
We compare to SDC with fixed iteration number and step size as a reference with no special resilience capabilities, which we call the ``fixed'' scheme.
Before presenting how the strategies perform for \replace{all problems from~autoref{sec:Problems}}{the PDE benchmarks}, we investigate them in detail with the Lorenz attractor problem.
\autoref{fig:fault_bit_0} and~\autoref{fig:fault_bit_20} illustrate for two specific experiments how susceptible this problem is to small perturbations.

\begin{figure}
    \centering
    \includegraphics{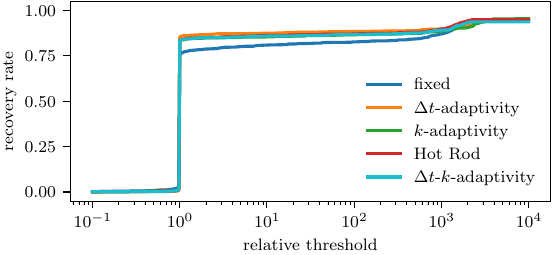}
    \caption{Recovery rate for the Lorenz attractor problem depending on the acceptance threshold.
    The threshold is shown as \replace{a ratio to the fault-free error}{the ratio of the global error to the global error of a fault-free run}, meaning a value of one requires fault correction to deliver exactly the same error as a simulation without faults. A value of two would count twice the error as without faults as recovered.
    Any threshold larger than one leads to high recovery rates for all recovery strategies.
    We choose a threshold of 1.1 for computing recovery rates.
    }
    \label{fig:recovery_rate_per_thresh}
\end{figure}

We select a recovery threshold that controls how much the global error is allowed to be increased relative to a fault-free run and compute the recovery rate as the ratio of experiments where the global error did not increase beyond the threshold to total number of experiments.
We count a fault as recovered when the global error is not increased by more than 10\,\%.
\autoref{fig:recovery_rate_per_thresh} shows how recovery rates change for different acceptance thresholds.
\deleted{In practice, values between one and two are probably reasonable.}
We find that in the case of the chaotic Lorenz attractor the exact value within this range does not matter much.
\added{Note that the acceptable size of perturbations due to faults is dependent on the application and the value of 10\,\% may not be suitable for all users.}

\begin{figure}
    \centering
    \includegraphics{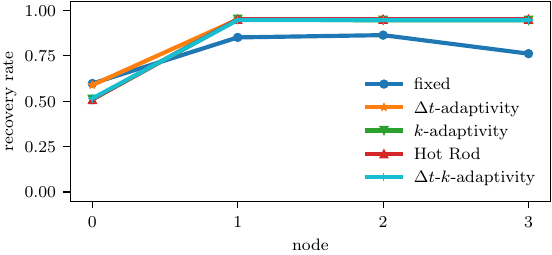}
    \caption{
    Recovery rate for the Lorenz attractor problem depending on the collocation node that was hit by a fault.
    As node 0 stores the initial conditions in the pySDC implementation, faults targeting it can not be recovered by restarting, which leads to decreased recovery rate compared to other nodes.
    }
    \label{fig:recovery_rate_Lorenz_node}
\end{figure}

\autoref{fig:recovery_rate_Lorenz_node} shows recovery rates dependent on the collocation node that is targeted by the fault.
Note that in the pySDC implementation, collocation node 0 stores the initial conditions.
This means faults that \replace{hit}{occur} in collocation node 0 cannot be recovered by restarting, since the initial conditions will continue to include the fault.
This causes significantly lower recovery rate for faults in node 0 than in other nodes.
We also find that the fixed strategy has slightly lower recovery rate in node 3 than in nodes 1 and 2.
After SDC has converged, the solution to the step is computed as a weighted sum of the $u_m$ during the so called collocation update.
With the Gau{\ss}-Radau quadrature rule with three nodes that we use here, the last node is the end point of the interval and the collocation update amounts to simply copying $u_3$.
Hence, when a fault targets any but the last node in the last iteration, it has no impact on the solution to the step.

\begin{figure}
    \centering
    \includegraphics{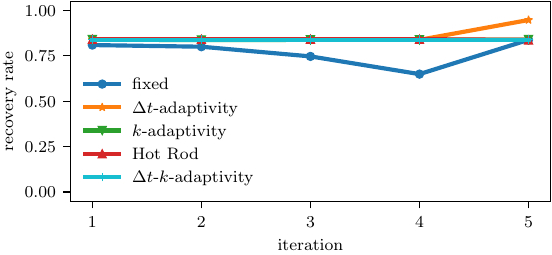}
    \caption{
    Recovery rate for the Lorenz attractor problem depending on the iteration in which the fault occurred.
    With the fixed strategy, the earlier the fault happens, the more likely it is to be smoothed out.
    This causes the recovery rate to decrease with iteration number.
    The exception to this rule is the last iteration, where only faults to the last node have an impact on the solution.
    As virtually all faults that do not target node 0 are fixed with all other strategies, we find no dependence of recovery rate on the iteration in which it occurs.
    Again, the last iteration is an exception for $\Delta t$-adaptivity, because faults that target the initial condition have no impact with fixed iteration number.
    }
    \label{fig:recovery_rate_Lorenz_iteration}
\end{figure}

\autoref{fig:recovery_rate_Lorenz_iteration} shows recovery rates depending on the iteration in which the fault occurs.
Since faults in the last iteration and any but the last node do not affect the collocation update, the fixed and $\Delta t$-adaptivity strategies have the highest recovery rate in the last iteration.
Because the fault nevertheless impacts the residual, strategies with adaptive $k$ may continue to iterate and propagate the fault into the collocation update.
Keep in mind that Hot Rod performs one extra iteration, but advances in time with the solution of the fifth iteration.
If a fault occurs in the initial conditions in the fifth iteration, Hot Rod will trigger a restart and the fault ends up affecting the solution.
Otherwise, the recovery rate of the fixed strategy decreases with iteration number.
The earlier the fault occurs, the more iterations are available to smooth out the perturbation.

\begin{figure}
    \centering
    \includegraphics{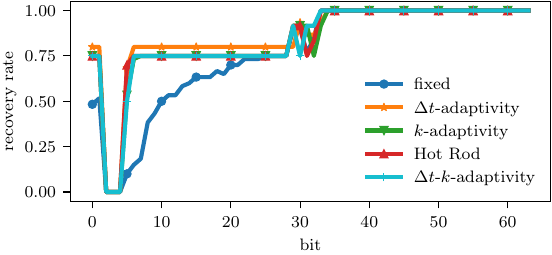}
    \caption{
    Recovery rate for the Lorenz attractor depending on the bit that is flipped by the fault.
    In this IEEE 754 64 bit floating point representation, bit 0 stores the sign, bits 1 through 12 store the exponent and the remaining bits store the mantissa.
    Flipping some bits in the exponent leads to overflow errors in the Newton solver and crash the code, resulting in 0\,\% recovery rate for any strategy.
    Flipping bits beyond 35 has insignificant impact on this simulation and does not require any recovery strategy.
    }
    \label{fig:recovery_rate_Lorenz_bit}
\end{figure}

Finally, we look at the recovery rate depending on which bit was flipped in \autoref{fig:recovery_rate_Lorenz_bit}.
In IEEE 754 double representation, bit 0 stores the sign, bits 1 to 12 store the exponent and the remaining bits store the mantissa.
We find that flipping bits beyond 35 perturbs the solution too little to be noticable in this problem.

When flipping bits 2, 3, or 4, overflow errors in the Newton solver crash the code, resulting in 0\,\% recovery rate for all strategies.
For $k$-adaptivity, $\Delta t$-$k$-adaptivity and Hot Rod strategies, we observe that 75\,\% of faults to bits 0, 1 and 9 to 28 are fixed.
This is because all faults that target these bits and not the initial conditions are fixed, while all faults that target these bits and the initial conditions are not fixed.
At 80\,\% the recovery rate for $\Delta t$-adaptivity is slightly higher in these bits because also faults that occur in the initial conditions and in the last iteration are fixed.

\begin{figure}
    \centering
    \includegraphics{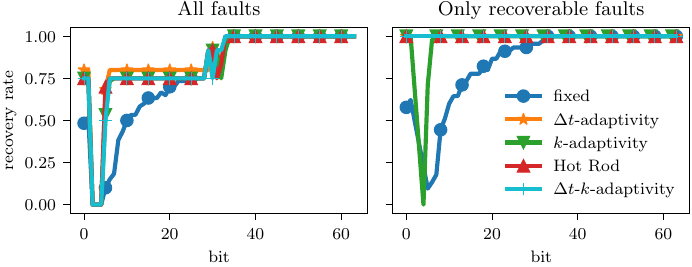}
    \caption{Recovery rate for the Lorenz attractor per bit. The left panel shows the recovery rate across all experiments, while the right panel shows only ones with faults that can be recovered. This means we exclude faults to the initial conditions in any but the last iteration and also faults that crash the code due to overflow errors in the Newton solver.}
    \label{fig:recoveryRate}
\end{figure}

After having identified the faults that cannot be recovered by these strategies as faults that target the initial conditions and faults that cause overflows, we can exclude these to compare the capabilities of the strategies amongst each other.
The recovery rate per bit with all faults and with only the ones that can be recovered is shown in~\autoref{fig:recoveryRate}.
All resilience strategies perform very well and correct nearly 100\,\% of the faults that they are theoretically able to.
Only $k$-adaptivity struggles with some faults to exponent bits because the Newton solver does not converge for arbitrary initial guess.
Strategies that include restarts are better equipped to deal with this type of fault.

\begin{figure}
    \centering
    \includegraphics{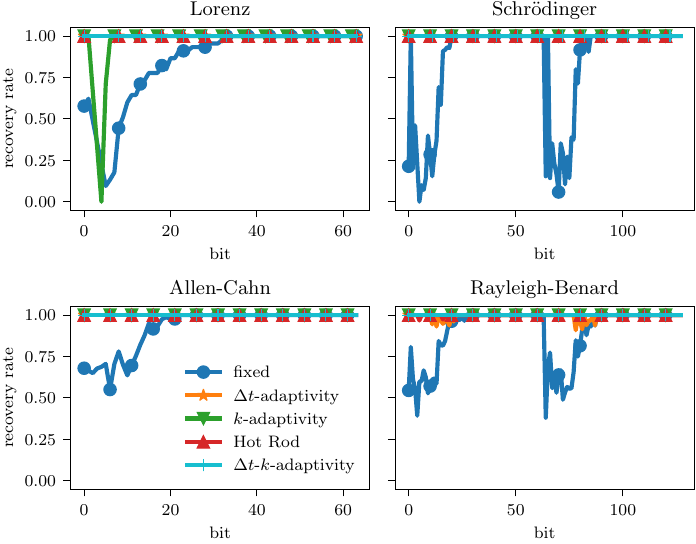}
    \caption{Comparison of recovery rate for theoretically recoverable faults for all problems from~\autoref{sec:Problems}. The Schr{\"o}dinger and Rayleigh-Benard problems use complex128 numbers instead of the usual float64, which we show as two float64 numbers back to back. The impact of faults appears to be the same per bit regardless of whether the complex or the real part is targeted.}
    \label{fig:recovery_compared_accross_problems}
\end{figure}

The resilience properties we identified for the Lorenz problem carry over to the PDE examples from~\replace{autoref{sec:Problems}}{Appendix~\ref{appendix:Problems}} as well.
We show the recovery rate for recoverable faults only in \autoref{fig:recovery_compared_accross_problems}.
For the Schr{\"o}dinger and Allen-Cahn problems, we find perfect recovery rates with all resilience strategies.
Since these problems are solved using an IMEX scheme which does not contain a Newton solver, $k$-adaptivity does not struggle with faults to exponent bits as is the case in the Lorenz attractor problem.
For RBC, we find a few faults in intermediate bits and late iterations that we cannot recover with $\Delta t$-adaptivity.
They do not trigger a restart, but end up noticeably changing the final solution.
Note that the absolute perturbation is small, but we solve the problem very accurately because the IMEX scheme is stable only for small step sizes.

Also, some faults are not detected by Hot Rod, which can be fixed by a tighter threshold.
We have selected a detection threshold such that no restart is triggered in the absence of faults, meaning no false positives.
In the original publication, the authors propose schemes for generating detection thresholds with different balances between allowing false positives and preventing false negatives~\cite{HotRod}.
Note that adaptive methods never exhibit false positives because the detection threshold is coupled to the accuracy of the problem.

\section{Discussion}
\label{sec:Discussion}
We demonstrate that adaptivity in SDC can be leveraged successfully as a resilience strategy against bit flips in part of the code.
Virtually all faults that can be recovered by these strategies are in fact recovered in the experiments.

We identified two classes of faults that require separate treatment.
First, some faults to exponent bits cause overflows and crash the code.
A simple remedy that can be added to all strategies is to attempt a restart whenever the Newton solver, or SDC, did not converge.
Second, faults to the initial conditions cannot be recovered by restarting from the same initial conditions.
Replicating the initial conditions only could protect against this kind of fault.
Another possibility would be to store the initial conditions in a part of memory that is especially well protected by hardware error correction codes, while the remaining data may be stored in memory with less or no error correction.

We compare against the dedicated resilience strategy Hot Rod for reference.
In our experiments, adaptive schemes were able to protect against faults almost perfectly, on the same level as Hot Rod.
It should be noted that the fault detection in Hot Rod is more sensitive in principle, which may be of more significant use in other applications.
However, Hot Rod adds substantial overhead both in computation and in memory, while adaptivity reduces computational overhead.
Not only does Hot Rod require additional resources, but the increased run time and memory footprint also increase the likelihood of a fault occurring in the simulation.

We conclude that SDC, equipped with step size adaptivity, is a very viable method that provides good computational efficiency and resilience against soft faults on par with a dedicated resilience strategy. 
In particular, the results also apply directly to the parallel-in-time algorithm diagonal SDC.

\section*{Acknowledgements}
We thankfully acknowledge funding from the European High-Performance Computing Joint Undertaking (JU) under grant agreement No 955701 and 101118139.
The JU receives support from the European Union’s Horizon 2020 research and innovation programme and Belgium, France, Germany, and Switzerland. 
We also thankfully acknowledge funding from the German Federal Ministry of Education and Research (BMBF) grant 16HPC048. 
The authors gratefully acknowledge computing time granted on JUSUF through the CSTMA project at J{\"u}lich Supercomputing Centre.
\added{The authors gratefully acknowledge the helpful comments from the reviewers.}

\section*{Availability of supporting data}
For instructions on how to reproduce the results shown in this paper, please \replace{consult the project section in the documentation of the pySDC code at https://parallel-in-time.org/pySDC/projects/Resilience.html}{refer to the instructions in \path{pySDC/projects/resilience/README.rst} in version 5.6 of pySDC~\cite{pySDC5_6}}.

\appendix
\section{Benchmark Problems}
\label{appendix:Problems}
\added{
Here, we describe the benchmark problems used in numerical experiments in detail.
We denote the simulation time at which we insert faults as $t_\mathrm{fault}$.
}

\subsection{Lorenz attractor \label{sec:LorenzProblem}}
\begin{figure}
    \centering
    \includegraphics{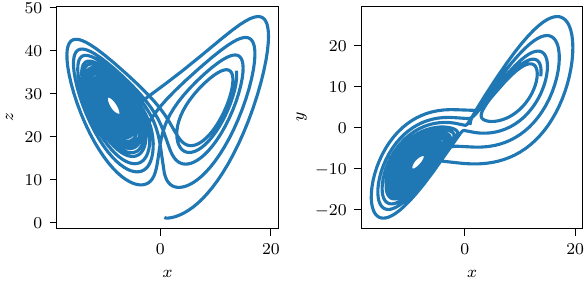}
    \caption{Solution of the Lorenz attractor problem over time. Shown are projections in the $x$-$z$ plane (left) and $x$-$y$ plane (right). The trajectory oscillates around two attractors in a chaotic manner. Faults are inserted when the solution is still circling the first attractor.}
    \label{fig:Lorenz-sol}
\end{figure}

Lorenz introduced this problem as a simplified system modelling the chaos that causes difficulty in numerical weather prediction \cite{lorenz_deterministic_1963}.
The problem traces its origins to an experiment where Lorenz wanted to show that numerical weather prediction is superior to statistical weather prediction.
He found that repeating a simulation yielded a different result after some simulation time than before, even though the calculations are deterministic.
He eventually realized that the cause was in rounding the solution to three digits when feeding it back in as initial conditions.
These small deviations would then continue to grow, as is characteristic of chaotic systems.
The problem is a non-linear coupled system of ODEs,
\begin{align}
    \begin{split}
    x_t & = \sigma \left(y - x\right), \\
    y_t & = \rho x - y - xz, \\
    z_t & = xy - \beta z,
    \label{eq:Lorenz}
    \end{split}
\end{align}
where we use parameters and initial conditions 
\begin{align}
    \begin{split}
    \sigma & = 10,~\rho = 28,~\beta = 8/3, \\
    x & (t=0) = y(t=0) = z(t=0) = 1, \\
    t & \in [0, 20],\\
    t_\mathrm{fault} &= 10,
    \end{split}
\end{align}
and is the result of Lorenz simplifying as much as possible while maintaining the key feature of growing perturbations~\cite{easterbrook_computing_2023}.
\autoref{fig:Lorenz-sol} shows the solution.
We \replace{solve the problem implicitly with}{integrate \autoref{eq:Lorenz} implicitly in the SDC iterations using} a self-constructed Newton solver and use the SciPy~\cite{SciPy} method \textsc{solve\_ivp} from the \textsc{integrate} package with an explicit embedded RKM of order 5(4)~\cite{DORMAND198019} with tolerances close to machine precision to obtain reference solutions.

\subsection{Non-linear Schr{\"o}dinger \label{sec:SchroedingerProblem}}
\begin{figure}
    \centering
    \includegraphics{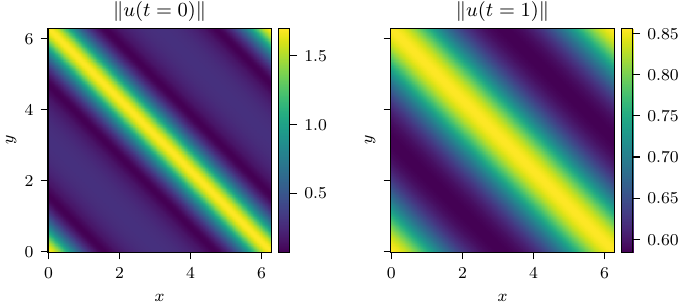}
    \caption{Initial condition for the Schr{\"o}dinger example (left) and the solution at the end of the simulation at $t=1$ (right). The initial condition is purely real, but the solution shifts to the complex domain as the simulation progresses. \added{Reproduced from~\cite{saupe2024adaptivetimestepselection} under CC~BY~4.0 licence.}}
    \label{fig:SchroedingerSolution}
\end{figure}

The focusing non-linear Schr{\"o}dinger equation is a wave-type equation that describes problems such as signal propagation in optical fibers~\cite{NLS_stuff}.
The formulation we solve can be written as
\begin{align}
    \label{eq:schroedinger}
    u_t =& i\Delta u + 2i \|u\|^2u,
\end{align}
with $i$ the imaginary unit.
We discretize with a Fourier based pseudo-spectral method and use implicit-explicit (IMEX) splitting to integrate the Laplacian implicitly and the non-linear term explicitly.
The domain and initial condition are
\begin{align}
    u(t=0) =& \frac{1}{\sqrt{2}} \left(\frac{1}{1 - \cos(x + y)/ \sqrt{2}} - 1\right), \\
    x &\in [0, 2\pi[^2, \\
    t &\in [0, 1], \\
    t_\mathrm{fault} &= 0.3,
\end{align}
with periodic boundary conditions.
The global error is computed with respect to the analytic solution~\cite[Equation (39)]{Schroedinger_exact_original}.
The solution is shown in \autoref{fig:SchroedingerSolution}.

\subsection{Allen-Cahn}
\label{sec:allen_cahn_problem}
\begin{figure}
    \centering
    \includegraphics{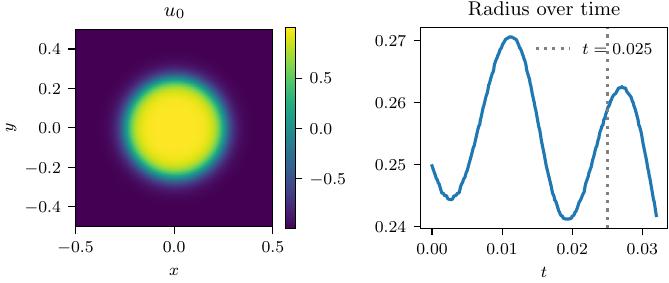}
    \caption{Left: Initial condition consisting of a circle of high phase embedded in the low phase. Right: Evolution of the radius of the circle over time computed from the analytical solution. The simulation runs until $t=0.025$. \added{Reproduced from~\cite{saupe2024adaptivetimestepselection} under CC~BY~4.0 licence.}}
    \label{fig:ACSolution}
\end{figure}

The variant of Allen-Cahn equation considered here is a two-dimensional reaction-diffusion equation with periodic boundary conditions that can be used to model transitions between two phases.
We choose initial conditions representing a circle of one phase embedded in the other phase.
We add time-dependent forcing, such that the circle alternates between growing and shrinking
\begin{align}
    u_t &= \Delta u - \frac{2}{\epsilon^2}u(1-u)(1-2u) - 6u(1-u)f(u, t),\\
    f(u, t) &= \frac{\sum \left(\Delta u - \frac{2}{\epsilon^2}u(1-u)(1-2u)\right)}{\sum 6u(1-u)}\left(1 - \sin{\left(4\pi \frac{t}{0.032}\right)}\times 10^{-2}\right).\\
\end{align}
The initial condition, parameters and domain are
\begin{align}
    u_0(x) &= \tanh{\left(\frac{R_0\|x\|}{\sqrt{2}\epsilon}\right)},\\
    \epsilon &= 0.04,\\
    R_0 &= 0.25, \\
    x &\in {[-0.5, 0.5[^2},\\
    t &\in [0, 0.025],\\
    t_\mathrm{fault} &= 0.01.
\end{align}
\autoref{fig:ACSolution} shows the initial condition and the evolution of the radius over time.
Similar to our approach for the Schr{\"o}dinger problem, we use IMEX splitting to integrate the Laplacian discretized with a Fourier spectral method implicitly while treating the non-linear term explicitly in physical space.
We again compute the error with respect to the SciPy method \textsc{solve\_ivp}.

\subsection{Rayleigh-Benard Convection}
\label{sec:RBC_problem}
Rayleigh-Benard convection (RBC) describes the dynamics that occur in a fluid that is heated from below, cooled from above, and subject to gravity.
The governing equations can be written as
\begin{align}
\begin{split}
        u_t - \nu (u_{xx} + u_{zz}) + p_x &= -uu_x - vu_z, \\
        v_t - \nu (v_{xx} + v_{zz}) + p_z - T &= -uv_x - vv_z, \\
        T_t - \kappa (T_{xx} + T_{zz}) &= -uT_x - vT_z, \\
        u_x + v_z &= 0,
        \label{eq:RBC}
\end{split}
\end{align}
with $u$ denoting the velocity in $x$-direction, $v$ the velocity in $z$-direction, $T$ the temperature, $p$ the pressure, \replace{and $\kappa=\sqrt{Ra \cdot Pr}$}{$\kappa$ the thermal diffusivity,} and \replace{$\nu = \sqrt{Ra / Pr}$, where $Ra=2\times 10^4$ and $Pr=1$ are the Rayleigh and Prandtl numbers.}{$\nu$ the kinematic viscosity.}
\added{
We use a Rayleigh number of
\begin{equation}
    Ra = \frac{\Delta T L_z^3}{\kappa \nu} = 3.2\times 10^5,
    \label{eq:RayleighNumber}
\end{equation}
with $\Delta T=2$ the temperature difference between the top and bottom plates and $L_z=2$ the separation between the top and bottom plates, and a Prandtl number of 
\begin{equation}
    Pr=\frac{\nu}{\kappa} = 1.
\end{equation}
}
\reviewcomment{We had made a mistake in the definition of the Rayleigh number in the previous version and quoted it too low. The simulations remain the same.}
We choose a spatial domain of size $\Omega = [0, 8)\times (-1, +1)$, with $256\times 128$ degrees of freedom, periodic boundary conditions in $x$-direction, Dirichlet boundary conditions 
\begin{align}
\begin{split}
  u(z=-1) &= u(z=1) = v(z=-1) = v(z=1) = T(z=1) = 0\\
  T(z=-1) &= 2,
\end{split}
\end{align}
in $z$-direction and $\int_\Omega p = 0$ for the pressure gauge.
We discretize the problem with a pseudo-spectral method using a Fourier base \replace{horizontally}{in $x$-direction} and an ultraspherical base~\cite{UltrasphericalMethod} \replace{vertically}{in $z$-direction}.
We treat the linear parts, which are written on the left hand side of \autoref{eq:RBC}, implicitly in spectral space and the non-linear convective terms on the right hand side explicitly in physical space.

\begin{figure}
    \centering
    \includegraphics[width=\textwidth]{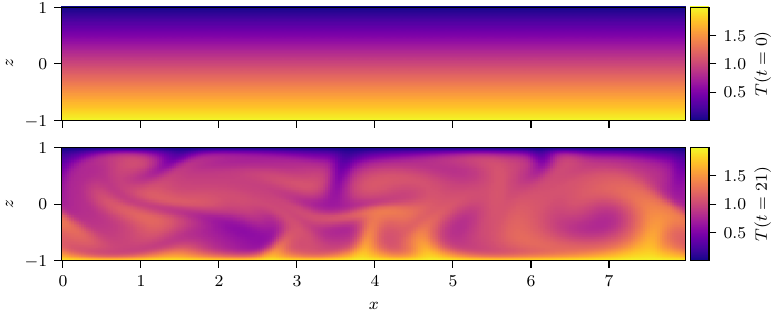}
    \caption{Temperature profile in the Rayleigh-Benard convection problem at the start and end of a simulation.
    The initial conditions include small perturbations not visible to the naked eye on the order of $10^{-3}$, which grow to complex flow patterns.
    In the experiments performed here, the simulation is started from a reference solution at $t=20$.
    }
    \label{fig:RBCSolution}
\end{figure}

The initial conditions are zero in all quantities but the temperature, where we choose a vertical linear gradient and perturb randomly.
The temperature gradient produced by heating the bottom leads to a density gradient, which causes upwelling.
The Rayleigh number is a non-dimensionalized quantity describing the balance of gravitational forcing and viscous damping.
The larger the Rayleigh number, the larger the impact of gravity, leading to more complex flow patterns.
The Rayleigh number we use here is large enough to cause turbulent flow, but low enough not to require very high spatial resolution.
The resulting dynamics show some rising plumes that move to the side and are pushed back down after they reach the top.
This eventually leads to a pseudo-stationary profile with circular currents.
The experiments we do here start at $t=20$ and end at $t=21$ with faults inserted at $t_\mathrm{fault}=20.2$.
This is when the first plumes have already been pushed down again but the circular motion is just developing.
See~\autoref{fig:RBCSolution} for a plot of the initial condition and solution at $t=21$.

We generate a reference solution by using a low tolerance of $\epsilon_\mathrm{TOL}=10^{-8}$ in $\Delta t$-adaptivity.
We verified the reference solution by confirming the order of accuracy of converged collocation problems.
As the resolution is large, we restrict fault insertion to the 16 slowest modes in either direction to cause noticeable perturbations due to the faults.

\bibliography{bibliography.bib}
\bibliographystyle{spmpsci}


\end{document}